# An economic perspective on major cloud computing providers

Noman Islam
Associate Professor,
Iqra University
Karachi, Pakistan

noman.islam@gmail.com

Zeeshan Islam
Research Fellow,
ALADIN Solutions
Karachi, Pakistan

*Abstract*— Cloud computing can be defined as the outsourcing / renting of resources over the network. It has been regarded as an essential utility similar to electricity, water, telephone and sewerage utilities. Over the years, a number of providers have emerged that provide basic and advanced form of services for cloud computing. However, these providers employ their own nomenclature for marketing their specification and the range of services offered by them. This makes it very difficult for a common user to select a platform based on their requirements. This paper provides a comparison of different cloud service providers from an economic perspective. It performs the cost analysis of services provided by providers for different scenarios. Based on the analysis, conclusions have been drawn and directions for further research have been provided.

## I. INTRODUCTION

Due to the innovation in technologies, the past few years have witnessed the emergence of a large number of computing models. *Cloud computing* has also risen as a new model recently in which a collection of resources are dynamic provisioned to satisfy end user's requisition. It enables the end-user to perform their domestic jobs over the internet and can be charged as per the resource consumption. Different types of resources that can be offered over the cloud can be computing, storage, database and network [9]. The cloud can be deployed in public, private or as a community [10]. Different types of service models such as Infrastructure-as-a-Service (IaaS), Platform-as-a-Service (PaaS) and Software-as-a-Service (SaaS) can be supported [11]. The topic of cloud computing has been under intense investigation over the past few years and a number of research issues are being deliberated. Some of these research issues are security, privacy, energy management, virtualization and data management [12-14].

Over the years, different cloud computing service providers have emerged [1]. These providers provide various deployment schemes and service models. However, they employ their own nomenclature for their specification. This makes it cumbersome for a common person to make a decision for the selection of a provider based on their requirements. In this paper, a comparative study of major cloud computing providers has been provided from an economic perspective.

Rest of the section embarks upon the comparative study. The next section starts with an overview of recent literature related to the topic and builds the research question. The framework for comparison is provided afterwards. This is followed with the comparative analysis. The paper concludes with the summary of the research work and guidelines for future work.

## II. RELATED WORK

The research on cloud computing can be considered in infancy. Only recently, the venture capitalist has started to consider it as an investment option. This section deliberates on the literature on cloud from two perspectives. First, an overview of major comparative studies reported on cloud computing providers is provided. Then, some of the work on economic aspect of cloud computing is discussed.

### A. Comparative Studies

As far as the comparative studies are concerned, only a little amount of literature is available. A comparative study of different cloud platforms has been provided from marketing perspectives in [1]. The authors concluded that existing providers don't accord with the marketing demands. In [9], an analysis has been performed between four IaaS and PaaS providers. It is established that no current service providers are self-sufficient in meeting the varied demands of end-users. A qualitative study of different commercial cloud platforms have been provided in [4]. Different IaaS and PaaS providers have also been contrasted out in [2]. Some studies presented a tree-based taxonomy highlighting the characteristics of different service providers [5]. CloudComp employs an end-to-end performance comparison of different service providers [3]. A quantitative comparison of Amazon EC2 for scientific computing purpose has been provided in [7]. Different cloud providers have been compared based on their application models in [15]. Finally, a framework for ranking cloud computing providers has been provided in [16]. The

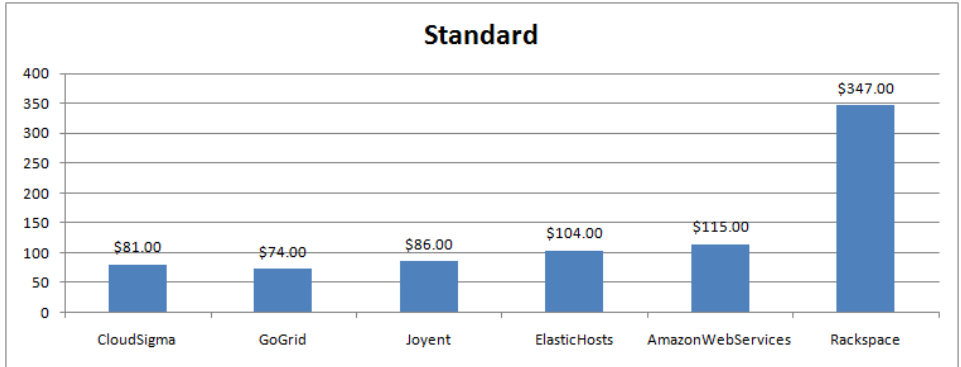
a. Cost of standard instance

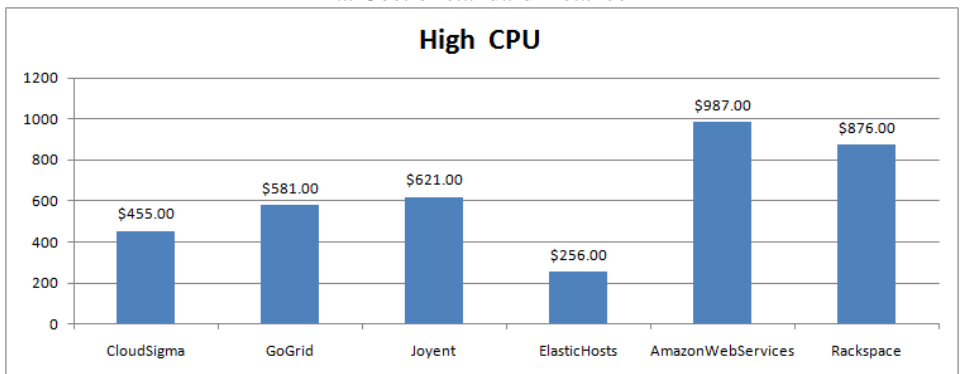
b. Cost of high CPU instance

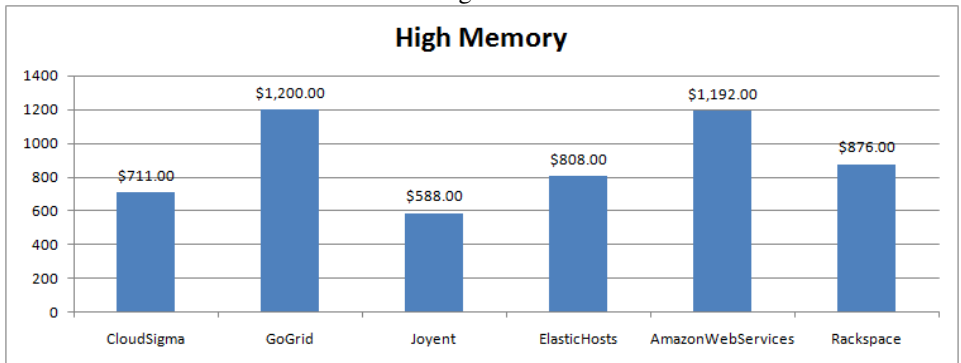
c. Cost of high memory instance

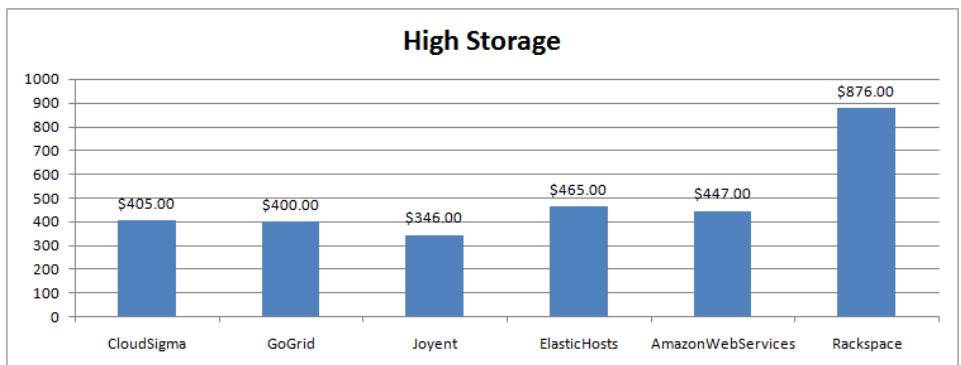
d. Cost of high storage instance

**Figure 1: Cost analysis of different cloud computing providers**

framework ranks the providers based on compliance with QoS requirements of end-users.

Based on the review of literature, it has been found that no work has been reported that compares the provider based on economic perspective.

### B. Economic perspective

There has been only a little work done on analyzing the economic aspect of cloud provider [22]. According to [19], cloud economics have the widest impact from business perspective. Discussions on economics of hybrid cloud have been provided in [20]. In [17], the pricing models of different cloud provider are compared and the pros and cons of each of the approaches are highlighted. Different types of pricing models exist such as fixed pricing and pay-per-use. [18] used the TOE framework for examining, among other things, the cloud computing pricing models. In [21], a study has performed for the impact of price competition and congestion in cloud market.

## III. COST ANALYSIS OF MAJOR CLOUD COMPUTING PROVIDERS

There have been a large number of cloud computing providers available in market, and according to some sources, the total number of providers are more than 80 cloud [8]. In our study, we restrict our-self to following major cloud providers currently available in market: CloudSigma, GoGrid, Joyent, ElasticHosts, Amazon EC2 and Rackspace.

### A. Benchmark Instances

To analyze the cost of providers, we consider four different types of instances corresponding to different scenarios. Table 1 lists down the benchmark instances used for comparison. The paper first analyzes the cost of different service providers for standard instance i.e. situation in which the end-user's requirements for memory, CPU and storage is moderate. Then, the paper analyzes the scenario in which CPU requirement is very high and identifies the cost of different service providers. Similarly, the service providers' cost are analyzed for high memory and high storage scenarios.

**Table 1: Details of benchmark instances for analysis**

| Instance name | Specifications | | |
|---|---|---|---|
| | RAM (GB) | CPU (vCPU) | Storage (GB) |
| Standard | 2 | 1 | 100 |
| High CPU | 8 | 8 | 100 |
| High Memory | 32 | 4 | 100 |
| High Storage | 8 | 4 | 1024 |

Service duration = 1 month, operating system = Linux

### B. A cost analysis of providers

Figure 1 analyses the cost of running the benchmark instances on different cloud computing providers. Let us analyze the cost of the providers for each particular instance.

**Standard instance:** Figure 1a analyzes the cost of executing a standard instance (for normal computing tasks) on different cloud providers. The *GoGrid* has the minimal cost offering for running a standard instance. It provides real-time provisioning and scaling of CPU, storage and network facilities in robust and resilient manner using its web control or API. For the standard scenario, GoGrid provides the following configuration:

Memory: 2 GB RAM,
Processor: 2x vCPU
Storage: 100 GB storage
Cost: $74

The platform maintains images for different 32 and 64 bit Windows and Linux operating systems. GoGrid uses Redundant Array of Independent Disk (RAID) 6+0 to provide resilience and provides100% guaranteed network delivery. A variety of SQL and NoSQL database solutions are provided.

Besides GoGrid, *CloudSigma* and *Joyent* also provides the next best offering in terms of cost for executing standard instance.

**High CPU instance:** Figure 1b shows the cost for a high CPU instance. The high CPU instances are useful for performing computationally intensive tasks. It can be seen that *Elastic Hosts* provides the least cost offering of $256 for the execution of a high CPU instance. It provides fast, flexible and easy to deploy cloud computing solution with its data centers all over the world. The provider offers an easy to use control panel, high throughput storage, preinstalled images for most of the operating systems and technical support over emails and phones.

It is to be noted that for *Rackspace*, we don't find any precise matching instance. Hence, the closest matching configuration is selected, which is as follows:

Memory: 30GB memory
Processor: 8 vCPU processor
Storage: 1.2TB storage
Cost: $876

**High memory instance:** Figure 1c illustrates the cost of executing a high memory instance. For high memory applications, *Joyent* offers the best price, while *CloudSigma* and *ElasticHosts* also provide optimal costs. Joyent provides compute, storage and private cloud services into end users. The compute service enables reliable and resilient applications on efficient and highly available cloud platform. The environment offers features like zoning, high integrity caching and tracing features etc. Joyent provides Manta service a highly scalable and distributed object storage service.

For *GoGrid* and *Rackspace*, no matching configuration is found, and the cost of closest configuration has been reported.

**High storage instance:** These types of instances are useful for applications that require large storage. Figure 1d analyzes the cost of different providers for these types of instances. It can be observed that *Joyent* provides the optimal cost for execution of high storage instances.

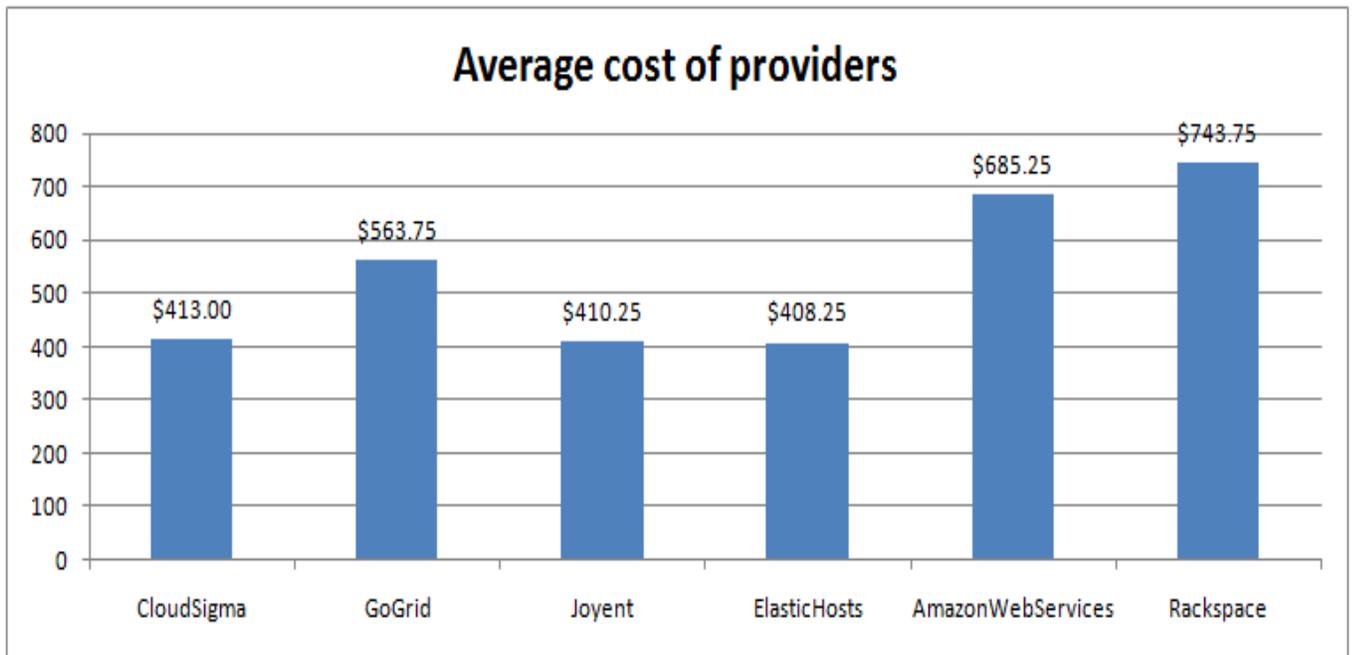

Figure 2: Average cost of providers for executing any type of instance

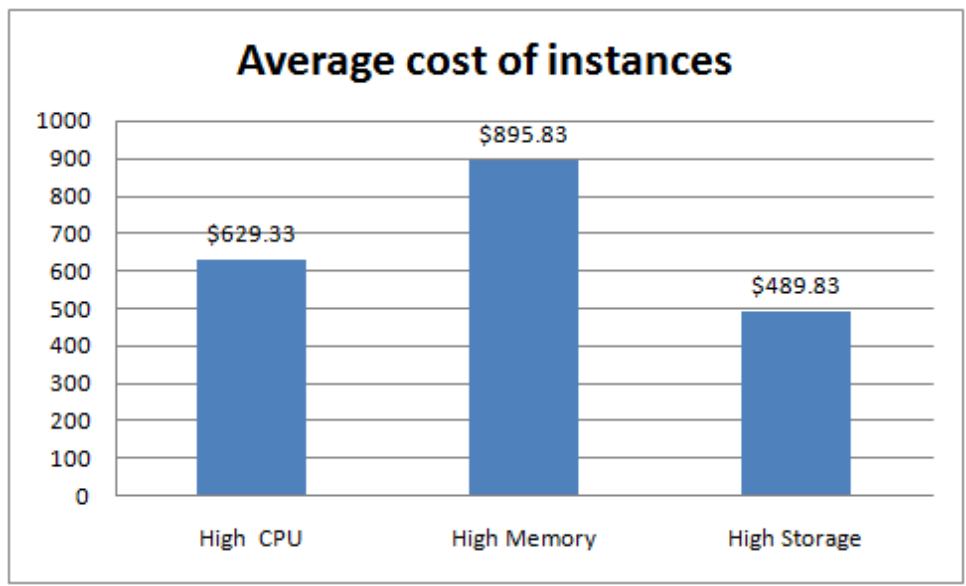

Figure 3: Average cost of instances

Figure 2 shows the average cost of executing any type of instance on different providers. It can be observed that CloudSigma, Joyent and ElasticHosts are economically most feasible, and Rackspace and AmazonWebServices are amongs the most expensive platforms for the configuration mentioned in Table 1. However, the additional cost associated with these platforms can be due to additional features provided by them, and require further investigation. For example, *Rackspace* provides features of scheduled imaging, cloud backup and snapshots etc. that improves the reliability of the system.

Figure 3 shows the overall average cost of different instances. The cost of running a high memory instance is higher. Hence, it can be concluded that major cost of cloud computing system is dominated by high memory instances.

ACKNOWLEDGMENT

The authors would like to thank the Iqra University, Karachi, Center for Research in Ubiquitous Computing, Karachi and ALADIN Solutions for providing support in the completion of this research work.

IV. CONCLUSION

This paper provides an economic perspectives on different cloud computing. A cost analysis of different providers is first provided. It has been concluded that CloudSigma, Joyent and ElasticHosts offer the best price for the selected benchmark instances. However, the analysis performed in this paper should be generalized further keeping into considering other parameters like network bandwidth, availability, reliability and other features of cloud providers. In addition, future work should also be done on analyzing these systems for other types of instances.